# The role of single oxygen or metal induced defect and correlated multiple defects in the formation of conducting filaments

Z. Zhang[†], H. Li[†*], L. Shi

Department of Precision Instrument, Centre for Brain Inspired Computing Research, Tsinghua University, Beijing, 100084

[†] *These authors contribute equally.*
[*] *Corresponding author email: li_huanglong@mail.tsinghua.edu.cn*

**Abstract**

We study the dependence of the formation energies of oxygen and metal induced defects in $Ta_2O_5$, $TaO_2$, $TaO$, $TiO_2$ and $Ti_4O_7$ on the chemical potential of electron and atomic constitutes. In the study of single defect, metal induced defects are found to be preferable to oxygen induced defects. This is against the experimental fact of the dominant role of oxygen induced defects in the RS process. A simple multiple defects picture without correlated atomic rearrangement does not cure this problem. The problem is resolved under the correlated multiple defect picture where the multiple defects result in correlated atomic rearrangement and the final products show certain atomic ordering.

**Introduction**

Resistive switching (RS) in transition metal oxides (TMOs) has attracted great interests for a possible application in nonvolatile memory devices [1, 2] and analog memristors [3-5]. Defect redistribution under the switching field is generally considered to play prominent role in the reversible RS [6-8]. In simple binary TMOs such as $Ta_2O_5$ and $TiO_2$, the agglomeration (segregation) of oxygen vacancies and the subsequent growth (rupture) of the conducting filament is one of the most cited RS mechanisms [6-8]. These simple binary TMOs have the advantage of easy fabrication. However, the role of oxygen vacancies in the formation of conducting filaments at an atomic level is less well known. Besides the locally formed conducting path, the resistance can also be switched back and forth by the drift of the charged oxygen vacancies towards or away from one of the two electrodes which modifies the width of the interfacial depletion region and consequently the Schottky contact resistance [9]. Oxygen vacancy induced charge trap can also assist bulk like RS through continuous filling or emptying the traps, whose extent of charge occupation varies the resistance [10]. More complicated correlated TMOs like $Pr_{0.7}Ca_{0.3}MnO_3$ [11], $SmNiO_3$ [12] show metal-insulator transition which accounts for the RS. In these correlated TMOs, the oxygen vacancy coordination environment, which is subject to the driving field either electrically or thermally, modulates the local valence state of transition metal by rearranging the hybridization of transition metal d and oxygen p orbitals. Above mentioned TMOs show RS through effective change of the local oxidation state. Therefore, they are categorized as Redox (Reduction-oxidation) RS TMOs [5]. In this work, we focus on the filament type RS. High performance $Ta_2O_5$ based nonvolatile memory device [13] and $TiO_2$ memristor based neuromorphic network [14] have recently been demonstrated. In situ characterization revealed that the conducting filaments in $TiO_2$ and $Ta_2O_5$ are composed of Magneli phase $Ti_4O_7$ [15] and $TaO_{1-x}$ [16], respectively. This hypo-stoichiometry can result from either oxygen vacancy or metal interstitial, as expressed in the Kroger–Vink notation by the following two Redox reactions, respectively.

$$O_O^X \rightarrow V_O^{\bullet\bullet} + 2e^{'} + \frac{1}{2}O_{2(g)}$$

and

$$M_M^X + O_O^X \rightarrow M_i^{\bullet\bullet} + 2e^{'} + \frac{1}{2}O_{2(g)}$$

Though the controversy in identifying the major defect type that leads to such non-stoichiometry exists, supporting evidence for oxygen vacancy as the major type of defect in the formation of conducting filaments has been reported [17]. Nevertheless, a comparative study of the behaviors of oxygen vacancies and metal interstitials in the formation of nanofilaments at an atomic level is useful. In this work, we aim to explain the reason for the dominant role of oxygen vacancies over metal interstitials in the formation of nanofilaments from a formation energy perspective. We find that a picture of correlated multiple defects, which is simply ignored in the previous studies, is needed to answer this question. By treating the defects as correlated, it is also possible to understand or even predict the morphology of the atomic structures of the nanofilaments.

**Computational methods**

This work is carried out by first principle calculation in the Cambridge Serial Total Energy Package (CASTEP) [18, 19]. The calculation of single defect is studied based on the supercell method, where the supercell size is large enough to mitigate spurious interaction with image defects and the defect can well be treated as isolated or single one. Given the large supercell size and only one defect, the lattices of the supercell can be fixed to those of the defect free one. The geometric structure is relaxed by GGA functional [20] with 380 eV cutoff energy of the plane wave basis set and ultrasoft pseudo-potentials. However, GGA is well known for underestimation of the band gap of semiconductors and insulators. Thus, we carry out screened exchange (sX-LDA) hybrid functional [21] to study the electronic structure of the GGA relaxed geometric structure. In sX-LDA calculation, 750 eV cutoff energy and norm conserving pseudopotentails are used. sX-LDA includes 100% exact exchange and takes the correlation effect into account by short ranged screening of the exchange potential. It accurately describes the electronic structures of various materials [22, 23] and has been widely used in the study of point defects [24-26]. In our study of uncorrelated multiple defects, multiple point defects are created in the pristine supercell without artificially rearranging the remaining atomic coordination. The lattice relaxation is allowed in this case. The correlation among multiple point defects, on the other hand, is treated by rearranging the defective system into certain morphology of the nanofilament. We use Magneli phase $Ti_4O_7$ and cubic TaO as the correlated atomic rearrangement of the filaments in $TiO_2$ and $Ta_2O_5$, respectively.

**Results and discussion**

**Bulk crystalline structures of main oxides**

The crystalline structures and sX-LDA band structures of $\lambda$-$Ta_2O_5$, rutile $TaO_2$, rutile $TiO_2$, Magneli phase $Ti_4O_7$ can be found in ref. [27] and those therein. Rutile phase $TiO_2$ is structural compatible with Magneli phase $Ti_4O_7$ and its switchable resistance is found [28] to occur in the absence of electro-forming process with high repeatability due to easy structure conversion to form the conducting filament, which is composed of Magneli phase $Ti_4O_7$ [15]. The morphology of nanofilaments in $Ta_2O_5$, on the other hand, is less well known. Recent in situ characterization showed the stoichiometry as $TaO_{1-x}$ and the evolution from short ranged order structure to long ranged ordering by the excess sweeping voltage and time [16]. Therefore, we model the structure of nanofilament in $Ta_2O_5$ by crystalline TaO. Reliable evidence for the existence of independent TaO is absent and it is observed in bulky sample only in the presence of higher oxide [29]. Micro-diffraction studies proposed various tentative TaO structures [31-34]. We adopt Harris et al [33] model, as shown in fig. 1(a). We find this model the only one immune to net atomic reconstruction due to single point defect, compared with the instability (not shown) of other model structures [30-32]. The sX-LDA band structure of TaO is shown in fig. 1(b) and the metallic property of TaO is seen.

**Single defect**

The formalism we have adopted for the calculation of point defect formation energy follows the method of Lany and Zunger [34]. The charge states from -2 to +2 are studied for different point defects. The charge transition level diagram of single point defect in tantalum oxide systems is shown in fig. 2(a). For $\lambda$-$Ta_2O_5$, there are three different oxygen sites: interlayer 2-fold site, interlayer 3-fold site and intralayer 2-fold site. An oxygen vacancy at the intralayer 2-fold site has the lowest formation energy over all charge states. We can see from the diagram that a transition from +2 charged state to neutral state at a Fermi level just below the conduction band minimum (CBM), in agreement with Guo et al [35]. $Ta_2O_5/TaO_2$ bilayer structure is useful for controllable programming current and device stability that $TaO_2$ serves as metallic source of oxygen ions for the switching in the insulating $Ta_2O_5$ [36, 37]. Our calculation [27] has shown that the system Fermi energy of $Ta_2O_5/TaO_2$ bilayer structure locates just above the CBM of $Ta_2O_5$, which leaves the -2 charged single oxygen vacancy more stable than the +2 charged one. A more detailed discussion on this issue and the proposed way to restore the stabilized +2 charged state can be found in ref. [27]. The single metal interstitial in $\lambda$-$Ta_2O_5$ stabilizes at the +2 charged state over the entire range of Fermi level across the band gap. For rutile $TaO_2$, we align its bulk Fermi energy to the band edges of $Ta_2O_5$ based on the interfacial calculation [27]. $TaO_2$ is hypo stoichiometric, we consider oxygen interstitial and metal vacancy here. This is relevant to the vanishing of nanofilaments and restoring the pristine $Ta_2O_5$. At the Fermi energy of $TaO_2$, the single oxygen interstitial is in the +2 charged state and the single metal vacancy is in the -2 charged state. For cubic TaO (not shown), the stable charge states of single oxygen vacancy and single metal defect are -1 and -2, respectively. The charge transition levels of the single oxygen vacancy from +2 charged to neutral, then to -1 charged cluster in a very narrow range of 0.02 eV, about 0.2 eV below its intrinsic Fermi energy, and no near lying charge transition level to the Fermi energy is found.

The charge transition level diagram of single point defect in titanium oxide systems is shown in fig. 2(b). For rutile $TiO_2$, the charge transition of oxygen vacancy from +2 charged state to neutral state, then to -2 charged states occurs in the close vicinity of the CBM of $TiO_2$, in agreement with recent GW result [38]. The calculation of $TiO_2/Ti_4O_7$ interface [27] also shows that the system Fermi energy is above the CBM of $TiO_2$. Similar to the tantalum system, the single oxygen vacancy in $TiO_2$ is stabilized at the -2 charged state according to the calculated system

Fermi energy of $TiO_2/Ti_4O_7$ interface [27]. The single metal interstitial is stabilized at the neutral state. For Magneli phase $Ti_4O_7$, the single oxygen interstitial is stabilized at the -1 charged state and the single metal vacancy is at the -2 charged state.

Although the charge transition levels of point defects do not vary under different oxygen chemical potentials (or the metal chemical potentials accordingly), the formation energies do. The chemical potentials are allowed to vary over a restricted range following the equilibrium thermodynamics. The boundary conditions of the chemical potentials are selected in the following way: in the study of bulky stoichiometric $TiO_2$, the chemical potentials satisfy $2\mu_O + \mu_{Ti} = H(TiO_2)$. As the reduction (filament formation) of $TiO_2$ terminates at the Magneli phase $Ti_4O_7$ so that we take the chemical potential of $TiO_2$ and $Ti_4O_7$ in thermal equilibrium as the O-poor condition. $\mu_O = \frac{1}{2}\mu_{O2(g)}$ is set as the O-rich condition, where $\mu_{O2(g)}$ is the chemical potential of single $O_2$ molecule. In the study of bulky hypo stoichiometric $Ti_4O_7$, the chemical potentials satisfy $7\mu_O + 4\mu_{Ti} = H(Ti_4O_7)$. As the $Ti_4O_7$ is oxidized (filament rupture) from thermal equilibrium with $TiO_2$ so that the O-rich condition is set to the chemical potential of $TiO_2$ and $Ti_4O_7$ in thermal equilibrium. The O-poor condition is set to $\mu_{Ti} = \mu_{Ti(hex)}$, where $\mu_{Ti(hex)}$ is the chemical potential of hexagonal crystalline Ti. For tantalum oxide systems ($Ta_2O_5$, $TaO_2$ and $TaO$), analogous selection rule for the boundary of the chemical potential applies.

Now, it is possible to study the formation energy of individual single point defect depending on the chemical potentials of the atomic constitutes as well as the electron chemical potential, namely the Fermi energy. Fig. 3(a) show the formation energies of single oxygen vacancy and single metal interstitial as functions of the oxygen chemical potential in $TiO_2$ at the system Fermi energy of $TiO_2/Ti_4O_7$. The charge states of the defects depend on the chosen Fermi energy and can be determined from fig. 2(b). The stable charge states of the defects may vary with the Fermi energy, the formation energies do not change significantly if we choose other characteristic values of Fermi energy in the vicinity of the tightly clustered charge transition levels. On the other hand, the formation energies do change significantly from the O-poor side to the O-rich side of the oxygen chemical potential. It is interesting to see that the single metal interstitial is energetically more favorable than the single oxygen vacancy on the O-poor side where the formation of Magneli phase $Ti_4O_7$ filament is most likely to occur.

Fig. 3(b) shows the formation energies of single oxygen interstitial and single metal vacancy as functions of the oxygen chemical potential in $Ti_4O_7$ at its intrinsic Fermi energy. It is clear that the single metal vacancy is more stable than the single oxygen interstitial over the entire range of oxygen chemical potential. In particular, at the O-rich condition for $Ti_4O_7$ (equivalent to the O-poor condition for $TiO_2$) where the filament starts to rupture the single metal vacancy is more stable by nearly 2.0 eV.

For $Ta_2O_5$, the formation energies of single oxygen vacancy and single metal interstitial as functions of the oxygen chemical potential at three characteristic values of Fermi energy are shown in Fig. 4(a). It is found that the single metal interstitial is more stable than the single oxygen vacancy on the O-poor side.

For $TaO_2$, the formation energies of single oxygen interstitial and single metal vacancy as functions of the oxygen chemical potential at its intrinsic Fermi energy are shown in Fig. 4(b). It is clear that the single metal vacancy is more stable than the single oxygen interstitial over the entire range of oxygen chemical potential. In particular, at the O-rich condition for $TaO_2$ (equivalent to the O-poor condition for $Ta_2O_5$) the single metal vacancy is more stable by nearly 2.0 eV.

Recent in situ study [16] observed that the filament stoichiometry in $Ta_2O_5/TaO_2$ is $TaO_{1-x}$, close to TaO. This is different than that in $TiO_2/Ti_4O_7$, where the filament stoichiometry is the same as that of the metallic oxygen vacancy reservoir. The selection rule for the boundary of oxygen chemical potential needs slight modification for the tantalum oxide system ($Ta_2O_5$ and TaO), as the O-poor condition for $Ta_2O_5$ should be determined from the thermal equilibrium of $Ta_2O_5$ and TaO, instead of $Ta_2O_5$ and $TaO_2$. This, however, only slightly changes the oxygen chemical potential at the O-poor condition and will not change the above statement regarding the relative stability of single oxygen vacancy and single metal interstitial in $Ta_2O_5$ at the O-poor side. The formation energies of single oxygen interstitial and single metal vacancy for TaO as functions of the oxygen chemical potential at the intrinsic Fermi energy are shown in Fig. 4(c). In contrast to other hypo stoichiometric oxides studied here, the single oxygen interstitial is more stable than the single metal vacancy.

The results of the oxides, except for TaO, based on the single point defect calculation all indicate that metal induced point defects are dominant in the formation and rupture of the nanofilaments, which is against the experimental evidence [17]. This is very likely due to the simplified single point defect picture, ignoring the fact that the defects are multiple.

## Multiple defects

Multiple point defects are created in the stoichiometric oxides by taking out several oxygen atoms or inserting several metal atoms so that the final stoichiometries are equal to those of the corresponding hypo stoichiometric oxides, modeling the formation of nanofilaments. Similarly for the hypo stoichiometric oxides, multiple point defects are created by inserting several oxygen atoms or taking out several metal atoms so that the final stoichiometries are equal to those of the corresponding stoichiometric oxides, modeling the rupture of nanofilaments. Due to much more number of point defects, both the internal atomic geometry and the supercell lattices are allowed to relax. It is worth pointing out that in building the model we only insert or take out atoms without additional rearrangement of other atoms in the host supercell. It is not easy to assign specific charge to such a number of individual point defects, we only consider all defects in the neutral state. The formation energies averaged to each point defect are shown in Fig. 5(a) for $Ta_2O_5$, where the O-poor condition is chosen at the thermal equilibrium between $Ta_2O_5$ and $TaO_2$. It is found that the averaged formation energy per point defect is larger than that of the single point defect. The averaged formation energy per oxygen vacancy is now lower than that of per metal interstitial at the O-poor side. In this sense, multiple defects picture does include additional effects that have been left out in the single defect picture, such as the interaction among point defects. For $TaO_2$, the formation energies averaged to each point defect are shown in Fig. 5(b). The averaged formation energy per point defect is larger than that of the single point defect and the metal vacancies are still more stable than the oxygen interstitials at the O-rich side. When the O-poor condition for $Ta_2O_5$ is chosen at the thermal equilibrium between $Ta_2O_5$ and TaO, the corresponding formation energy diagrams are shown in Fig. 5(c) and Fig. 5(d) for $Ta_2O_5$ and TaO, respectively. The averaged formation energy per oxygen vacancy is also found to be lower than that of per metal interstitial at the O-poor side for $Ta_2O_5$ in the multiple defects picture, in contrast to the single defect picture. For TaO, however, the averaged formation energy per oxygen interstitial is lower that of per metal vacancy at the O-rich side for TaO in both pictures. It is noteworthy that the formation energy difference is larger in the single defect picture than that in the multiple defect picture. For $TiO_2$, as shown in Fig. 5(e) the averaged formation energy per point defect is lower than that of the single point defect while the metal interstitials are still more stable than the oxygen vacancies at the O-poor side. For $Ti_4O_7$, as shown in Fig. 5(f) the averaged formation energy per point defect is larger than that of the single point defect and the metal vacancies are still more stable than the oxygen interstitials at the O-rich side.

The above results based on the multiple point defects picture show absolute values of the formation energy averaged to each point defect different than those based on the single point defect picture, indicating new effects that can only be presented by multiple defects. However, except for $Ta_2O_5$ and TaO, multiple defects picture still favors the metal induced defects over the oxygen induced ones in the formation and rupture of nanofilaments, which indicates that the multiple defects picture is still not enough. Actually, we have ignored the fact that the filament has certain ordered atomic structure but is not simply defective parent oxides. In particular, Magneli phase $Ti_4O_7$ [15] and certain long range ordered $TaO_{1-x}$ [16] are observed for titanium and tantalum systems, respectively. Besides, the ruptured filament merges into the $TiO_2$ or $Ta_2O_5$ matrix which should also has certain structure and cannot simply takes over the morphology of the filament. Therefore, it is important to take not only the multiple number of defects but also the correlated atomic arrangement into account. We name the full picture as the correlated multiple defects picture.

Recently, Xiao et al [39] varied the oxygen concentration in $TaO_x$ by randomly eliminating several oxygen atoms from amorphous $Ta_2O_5$ and found the tendency of forming Ta-Ta cluster. No correlated atomic arrangement was taken into account there. Park et al [38] modeled a chain of oxygen vacancies in $TiO_2$ and found that Ti ions nearby contribute the delocalized orbitals for electron conducting path. Kamiya et al [40,41] studied the cohesion (isolation) of the oxygen vacancies in the chain by charge injection (removal). However, the correlated atomic arrangement was oversimplified by using the chain model rather than the real Magneli phase $Ti_4O_7$ structure.

**Correlated multiple defects**

The formation energy averaged to each correlated multiple defect can be calculated simply by taking the aforementioned crystalline lower oxide as the reduced final product for filament formation, and crystalline stoichiometric oxide as the oxidized final product for filament rupture. As in the multiple defect picture, only neutral point defects are studied. Fig. 5(a), where the O-poor condition is chosen at the thermal equilibrium between $Ta_2O_5$ and $TaO_2$, compares the averaged formation energy per defect in multiple defect picture and correlated multiple defects picture. By taking the correlated atomic arrangement effect into consideration, the formation energies for both types of the defects are systematically lower than those calculated from the simple multiple defect picture. The oxygen vacancy is also found to be more stable at the O-poor side. For $TaO_2$ as shown in Fig. 5(b), the formation energies are systematically lowered by the correlated atomic arrangement effect and now the oxygen interstitials becomes more stable than the metal vacancies at the O-rich side. In Fig. 5(c), where the O-poor condition for $Ta_2O_5$ is chosen at the thermal equilibrium between $Ta_2O_5$ and TaO, similar effects of the correlated atomic arrangement are found as in Fig. 5(a). For TaO, however, the formation energy per oxygen interstitial is almost identical to that of per metal vacancy in the multiple correlated defects picture.

For TiO$_2$ as shown in Fig. 5(e), the formation energies are systematically lower by the correlated atomic arrangement effect than those calculated from the simple multiple defect picture. The oxygen vacancy is also found to be more stable at the O-poor side. For Ti$_4$O$_7$ as shown in Fig. 5(f), the formation energies are systematically lowered by the correlated atomic arrangement effect and now the oxygen interstitials becomes more stable than the metal vacancies at the O-rich side.

Liborio et al [42] have found the importance of the correlated atomic arrangement in the formation of the crystallographic shear planes in rutile. The above comparative results for titanium oxide system also demonstrate the importance of correlated multiple defects picture in understanding some phenomena in RS where numerous point defects are involved and interact with each other in a correlated manner. Given the significance of correlated atomic arrangement, the decrease of the relative stability of the oxygen vacancy over the metal interstitial in the sequence of 'single'>'multiple'>'correlated' and particularly the identical stability in the correlated multiple defects picture ('correlated') may be due to the real filament stoichiometry different than TaO and (or) the atomic ordering of the real filament in a more energetically stable structure other than the Harris one [33]. These differences can be subtle but significant enough to change the role of oxygen and metal induced defects in the RS process. For Ta$_2$O$_5$, the formation energy per oxygen vacancy is already smaller than that of per metal interstitial in the single defect picture. This is because single oxygen vacancy is already able to introduce some long ranged atomic rearrangement (not shown) to further stabilize the single oxygen vacancy, which has also been reported by Guo et al [44]. Such adaptive structure of Ta$_2$O$_5$ is useful for oxygen vacancy migration in the RS process.

**Conclusion**

We have studied the dependence of the formation energies of single point defect in Ta$_2$O$_5$, TaO$_2$, TaO, TiO$_2$ and Ti$_4$O$_7$ on both the chemical potential of the electron and the atomic constitutes. Oxygen vacancy and metal interstitial are considered in the stoichiometric oxides, and oxygen interstitials and metal vacancy in the hypo stoichiometric oxides, modelling the formation and rupture of the nanofilaments, respectively. According to this single defect picture, we find that the metal induced defects are more energetically preferable than the oxygen induced ones, which is against the experimental dominant role of the oxygen induced defects. By adopting a multiple defects picture, where the defective oxides with multiple defects but without obvious atomic rearrangement models the final products of the Redox process, no substantial improvement is found that the averaged formation energy per point defect still favors the metal induced defects. This controversy is resolved by taking the correlated atomic rearrangement into account. In this correlated multiple defects picture, the final products the Redox process are chosen to be oxides with certain atomic ordering according to the experiment, instead of being simple defective parent oxides. The oxygen induced defects are found to be dominant in this picture. The importance of taking the correlated atomic arrangement of multiple defects into account are demonstrated. It also provides a way to predict the morphology of nanofilaments under the restriction of thermodynamics.

**References**

1. H. S. P. Wong, H. Y. Lee, S. M. Yu, Y. S. Chen, Y. Wu, P. S. Chen, B. Lee, F. T. Chen, and M. J. Tsai, *Metal–Oxide RRAM*, Proc. IEEE 100, 1951 (2012).

2. H. S. P. Wong and S. Salahuddin, *Memory leads the way to better computing*, Nat. Nanotechnol. 10, 191 (2015).

3. J. J. Yang, D. B. Strukov, and D. R. Stewart, *Memristive devices for computing*, Nat. Nanotechnol. 8, 13 (2013).

4. D. Kuzum, S. Yu, and H.-S. P. Wong, *Synaptic electronics: materials, devices and applications*, Nanotechnology 24, 382001 (2013).

5. S. D. Ha and S. Ramanathan, *Adaptive oxide electronics: A review*, J. Appl. Phys. 110, 071101 (2011).

6. A. Sawa, *Resistive switching in transition metal oxides*, Mater. Today 11, 28 (2008).

7. R. Waser, R. Dittmann, G. Staikov, and K. Szot, *Redox-based resistive switching memories—Nanoionic mechanisms, prospects, and challenges*, Adv. Mater. 21, 2632 (2009).

8. K. M. Kim, D. S. Jeong, and C. S. Hwang, *Nanofilamentary resistive switching in binary oxide system; a review on the present status and outlook*, Nanotechnology 22, 254002 (2011).

9. J. J. Yang, M. D. Pickett, X. Li, D. A. A. Ohlberg, D. R. Stewart, and R. S. Williams, *Memristive switching mechanism for metal/oxide/metal nanodevices*, Nat. Nanotechnol. 3, 429 (2008).

10. J. H. Yoon, S. J. Song, I. H. Yoo, J. Y. Seok, K. J. Yoon, D. E. Kwon, T. H. Park, and C. S. Hwang, *Highly uniform, electroforming-free, and self-rectifying resistive memory in the Pt/Ta2O5/HfO2-x/TiN structure*, Adv. Funct. Mater. 24, 5086 (2014).

11. A. Odagawa, H. Sato, I. H. Inoue, H. Akoh, M. Kawasaki, Y. Tokura, T. Kanno, and H. Adachi, *Colossal electroresistance of a Pr0.7Ca0.3MnO3 thin film at room temperature*, Phys. Rev. B 70, 224403 (2004).

12. J. Shi, S. D. Ha, Y. Zhou, F. Schoofs, and S. Ramanathan, *A correlated nickelate synaptic transistor*, Nat. Commun. 4, 2676 (2013).

13. M. J. Lee, C. B. Lee, D. Lee, S. R. Lee, M. Chang, J. H. Hur, Y. B. Kim, C.-J. Kim, D. H. Seo, S. Seo, U.-I. Chung, I.-K. Yoo, and K. Kim, *A fast, high-endurance and scalable non-volatile memory device made from asymmetric Ta2O5-x/TaO2-x bilayer structures*, Nat. Mater. 10, 625–630 (2011).

14. M. Prezioso, F. Merrikh-Bayat, B. D. Hoskins, G. C. Adam, K. K. Likharev, and D. B. Strukov, *Training and operation of an integrated neuromorphic network based on metal-oxide memristors*, Nature 521, 61–64 (2015).

15. D.-H. Kwon, K. M. Kim, J. H. Jang, J. M. Jeon, M. H. Lee, G. H. Kim, X.-S. Li, G.-S. Park, B. Lee, S. Han, M. Kim, and C. S. Hwang, *Atomic structure of conducting nanofilaments in TiO2 resistive switching memory*, Nat. Nanotechnol. 5, 148 (2010).

16. G. S. Park, Y. B. Kim, S. Y. Park, X. S. Li, S. Heo, M. J. Lee, M. Chang, J. H. Kwon, M. Kim, U. Chung, R. Dittmann, R. Waser, and K. Kim, *In situ observation of filamentary conducting channels in an asymmetric TaO5-x/TaO2-x bilayer structure*, Nat. Commun. 4, 2382 (2013).

17. D. S. Jeong, H. Schroeder, U. Breuer, and R. Waser, *Characteristic electroforming behavior in Pt/TiO2/Pt resistive switching cells depending on atmosphere*, J. Appl. Phys. 104, 123716 (2008).

18. M.D. Segall, P.J.D. Lindan, M.J. Probert, C.J. Pickard, P.J. Hasnip, S.J. Clark, M.C. Payne, *First-principles simulation: ideas, illustrations and the CASTEP code*, J. Phys.: Condens. Matter.

14, 2717 (2002).

19. S.J. Clark, M.D. Segall, C.J. Pickard, P.J. Hasnip, M.J. Probert, K. Refson, M.C. Payne, *First principles methods using CASTEP*, Z. Kristallogr. 567, 220 (2005).

20. J. P. Perdew, K. Burke, and M. Ernzerhof, *Generalized Gradient Approximation Made Simple*, Phys. Rev. Lett. 78, 1396 (1997).

21. D. M. Bylander and L. Kleinman, *Good semiconductor band gaps with a modified local-density approximation*, Phys. Rev. B 41, 7868 (1990).

22. S. J. Clark, J. Robertson, *Screened exchange density functional applied to solids*, Phys. Rev. B, 82, 085208 (2010).

23. S. J. Clark and J. Robertson, *Calculation of semiconductor band structures and defects by the screened exchange density functional*, Phys. Status Solidi B, 248, 537 (2011).

24. S. J. Clark, J. Robertson, S. Lany, and A. Zunger, *Intrinsic defects in ZnO calculated by screened exchange and hybrid density functionals*, Phys. Rev. B 81, 115311 (2010).

25. J. Robertson and S. J. Clark, *Limits to doping in oxides*, Phys. Rev. B 83, 075205 (2011).

26. S. J. Clark and J. Robertson, *Energy levels of oxygen vacancies in BiFeO3 by screened exchange*, Appl. Phy. Lett. 94, 022902 (2009).

27. H. Li, Z. Zhang, and L. Shi, *The electronic structures of TiO2/Ti4O7, Ta2O5/TaO2 interfaces and the interfacial effects of dopants*, arXiv:1507.04479.

28. S. J. Song, J. Y. Seok, J. H. Yoon, K. M. Kim, G. H. Kim, M. H. Lee, and C. S. Hwang, *Real-time identification of the evolution of conducting nano-filaments in TiO2 thin film ReRAM*, Sci. Rep. 3, 3443 (2013).

29. A.I. Gusev, A.A. Rempel, and A.J. Magerl, *Disorder and Order in Strongly Nonstoichiometric Compounds: Transition Metal Carbides, Nitrides and Oxides* (Springer series in materials science; 47).

30. N. Norman, *Metallic Oxide Phases of Nb and Ta. I. X-Ray Investigation*, J. Less-Common Met., 4, 52 (1962).

31. V.I. Khitrova, *Electron Diffraction Study of Some Cubic Ta Oxides in Thin Films*, Sov. Phys. Crystallogr., I1(2), 199 (1966).

32. V.I. Khitrova and V.V. Klechkovskaya, *Electron Diffraction Investigation of the Thin Layers of Tetragonal Ta Oxides*, Sov. Phys. Crystallogr., 27(4), 441 (1982).

33. N. Harris, A. Taylor, and J. Stringer, *Kossel microdiffraction studies of TaO*, ActaMetall., 21, 1677 (1973).

34. S. Lany and A. Zunger, *Assessment of correction methods for the band-gap problem and for finite-size effects in supercell defect calculations: Case studies for ZnO and GaAs*, Phys. Rev. B 78, 235104 (2008).

35. Y. Guo and J. Robertson, *Materials selection for oxide-based resistive random access memories*, Appl. Phys. Lett. 105, 223516 (2014).

36. J. J. Yang, M.-X. Zhang, J. P. Strachan, F. Miao, M. D. Pickett, R. D. Kelley, G. Medeiros-Ribeiro, and R. S. Williams, *High switching endurance in TaOx memristive devices*, Appl. Phys. Lett. 97(23), 232102 (2010).

37. J. J. Yang and R. S. Williams, *Memristive Devices in Computing System: Promises and Challenges*, ACM J. Emerg. Technol. Comput. Syst. 9, 11 (2013).

38. A. Malashevich, M. Jain, and S. G. Louie, First-principles DFT + GW study of oxygen vacancies in rutile TiO2, Phys. Rev. B 89, 075205 (2014).

39. B. Xiao and S. Watanabe, *Oxygen vacancy effects on an amorphous-TaOx based resistance switch: a first principles study*, Nanoscale, 6, 10169 (2014).

40. S. G. Park, B. M. Kope, and Y. Nishi, *Impact of Oxygen Vacancy Ordering on the Formation of a Conductive Filament in TiO2 for Resistive Switching Memory*, IEEE Electron Device Lett. 32, 197 (2011).

41. K. Kamiya, M. Y. Yang, S. G. Park, B. M. Kope, Y. Nishi, M. Niwa, and K. Shiraishi, *ON-OFF switching mechanism of resistive–random–access–memories based on the formation and disruption of oxygen vacancy conducting channels*, Appl. Phys. Lett. 100, 073502 (2012).

42. K. Kamiya, M. Y. Yang, B. M. Kope, M. Niwa, Y. Nishi, and K. Shiraishi, *Vacancy Cohesion-Isolation Phase Transition Upon Charge Injection and Removal in Binary Oxide-Based RRAM Filamentary-Type Switching*, IEEE Trans. Electron Devices 60, 3400 (2013).

43. L. Liborio and N. Harrison, Thermodynamics of oxygen defective Magnéli phases in rutile: A first-principles study, Phys. Rev. B 77, 104104 (2008).

44. Y. Guo and J. Robertson, *Oxygen vacancy defects in Ta2O5 showing long-range atomic re-arrangements*, Appl. Phys. Lett. 104, 112906 (2014).

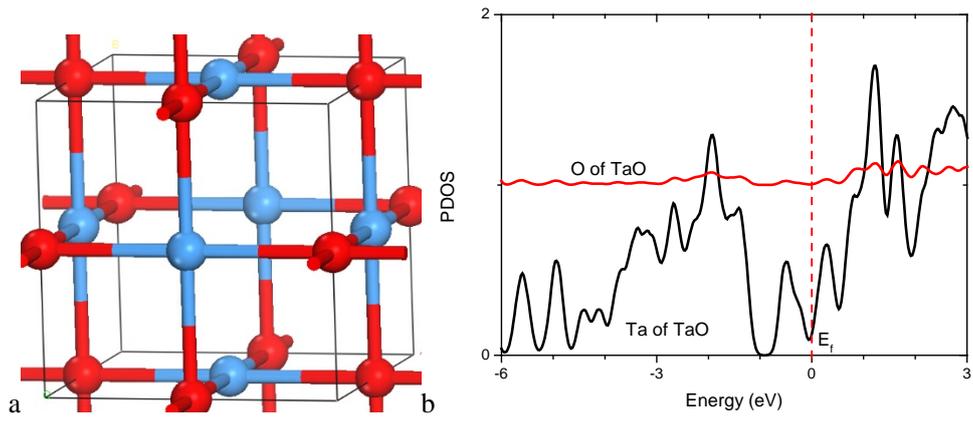

Fig. 1 (a) Atomic structure of cubic TaO model and (b) its sX-LDA band structure.

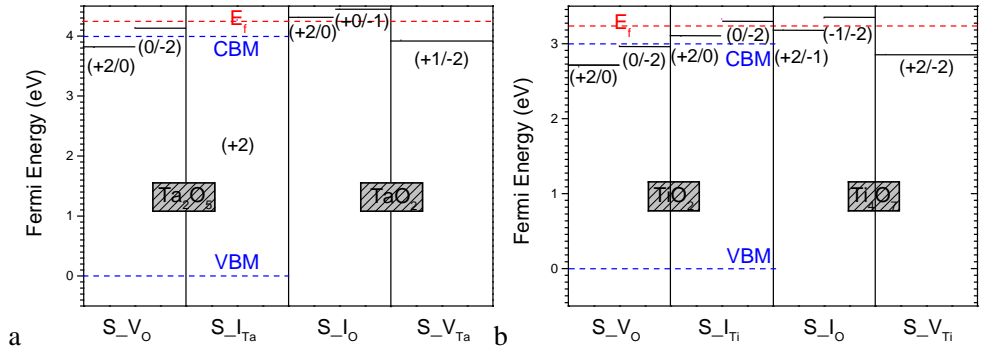

Fig. 2 The charge transition diagram for single oxygen vacancy (interstitial) and single metal interstitial (vacancy) in (a) $Ta_2O_5$ ($TaO_2$) and (b) $TiO_2$ ($Ti_4O_7$). The blue dotted lines show the band edges of the corresponding stoichiometric oxides, the red dotted lines show the system Fermi energies of $Ta_2O_5/TaO_2$ and $TiO_2/Ti_4O_7$, respectively.

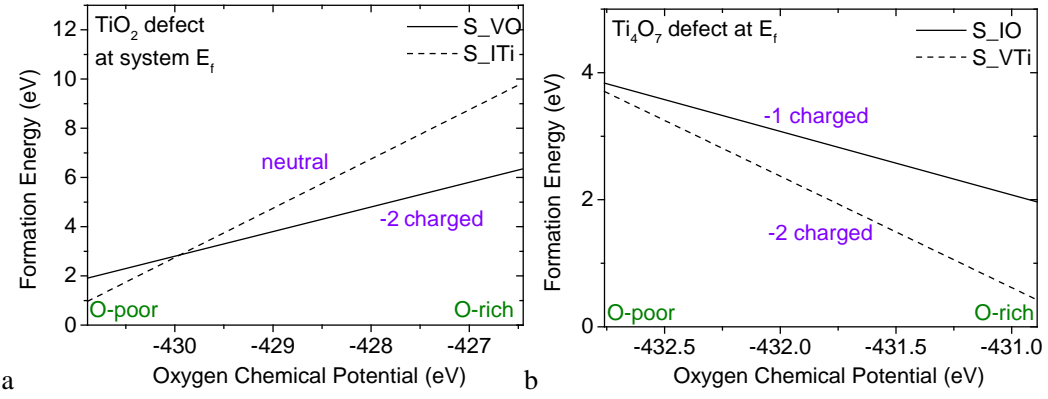

Fig. 3 (a) The formation energies of the single oxygen vacancy and single metal interstitial in $TiO_2$ as functions of the oxygen chemical potential; (b) The formation energies of the single oxygen interstitial and single metal vacancy in $Ti_4O_7$ as functions of the oxygen chemical potential. The electron chemical potentials are set to the system Fermi energy in both cases.

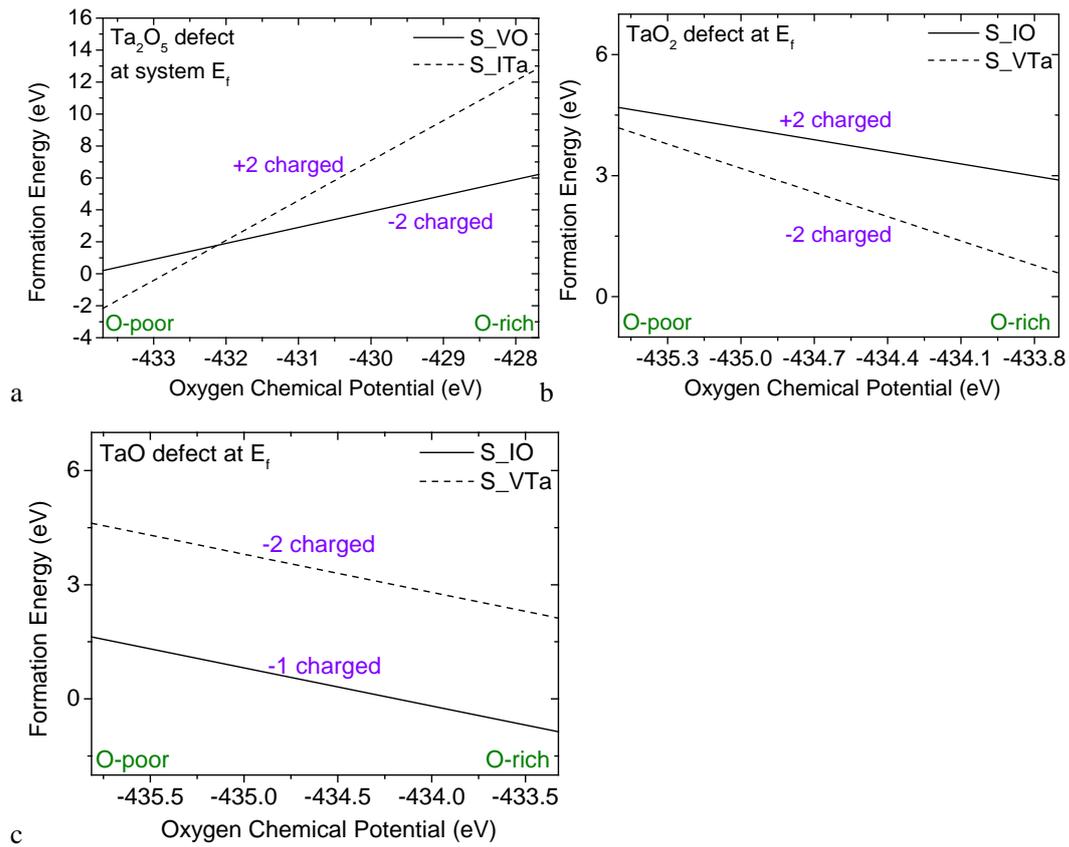

Fig. 4 (a) The formation energies of the single oxygen vacancy and single metal interstitial in $Ta_2O_5$ as functions of the oxygen chemical potential; (b) The formation energies of the single oxygen interstitial and single metal vacancy in $TaO_2$ as functions of the oxygen chemical potential; (c) The formation energies of the single oxygen interstitial and single metal vacancy in $TaO$ as functions of the oxygen chemical potential. The electron chemical potentials are set to the system Fermi energy in all cases.

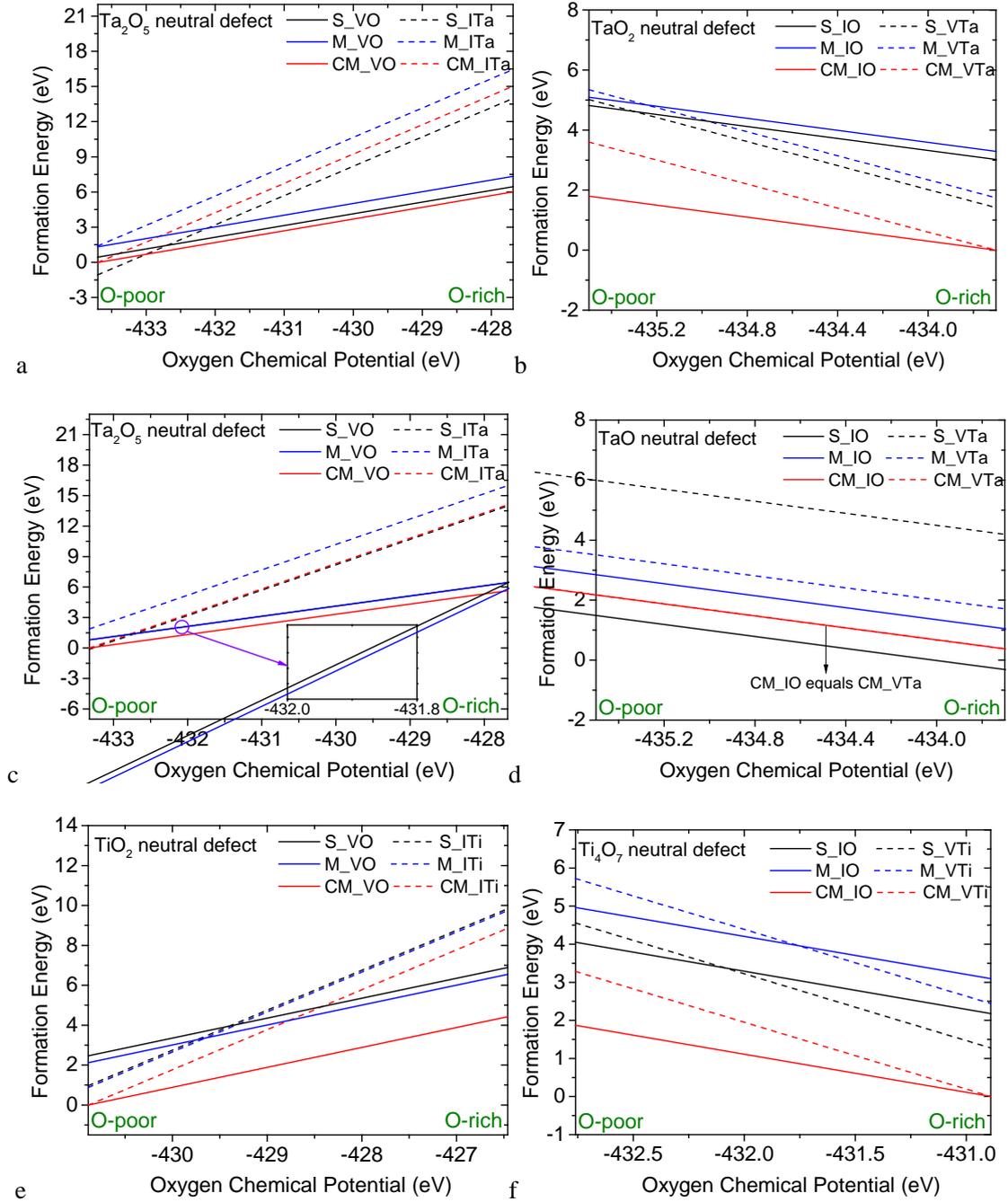

Fig. 5 The formation energies of the single/multiple/correlated multiple oxygen vacancy and single/multiple/correlated multiple metal interstitial in (a) $Ta_2O_5$, where $TaO_2$ is the result of the correlated atomic arrangement, (c) $Ta_2O_5$, where $TaO$ is the result of the correlated atomic arrangement, and (e) $TiO_2$ as functions of the oxygen chemical potential; the formation energies of the single/multiple/correlated multiple oxygen interstitial and single/multiple/correlated multiple metal vacancy in (b) $TaO_2$, (d) $TaO$ and (f) $Ti_4O_7$ as functions of the oxygen chemical potential. In all cases, only the neutral defect(s) is studied.